\documentclass[12pt]{article}
\usepackage{graphicx}

\newcommand{\text}{\rm}

\begin{document}

\title{\textbf{Remarks on the Gribov horizon and dynamical mass generation
in Euclidean Yang-Mills theories}\thanks{Work presented by R.F.
Sobreiro at IX Hadron Physics and VII Relativistic Aspects of
Nuclear Physics, Angra dos Reis, RJ, Brazil, March 28 to April 03,
2004.}}
\author{R.F. Sobreiro,
S.P. Sorella\thanks{sobreiro@uerj.br, sorella@uerj.br}\\
{\small{\textit{UERJ - Universidade do Estado do Rio de Janeiro,}}}\\
{\small{\textit{\ Rua S\~{a}o Francisco Xavier 524, 20550-013
Maracan\~{a}, }}}{\small {\textit{Rio de Janeiro, Brazil.}}} \and
D. Dudal,\
H. Verschelde\thanks{%
david.dudal@ugent.be,
henri.verschelde@ugent.be}\\
{\small {\textit{Ghent University}}}\\
{\small {\textit{Department of Mathematical Physics and Astronomy,
Krijgslaan 281-S9, }}}\\
{\small {\textit{B-9000 Gent, Belgium}}}}
\date{}
\maketitle

\begin{abstract}
The effect of the dynamical mass generation on the gluon and ghost
propagators in Euclidean Yang-Mills theory in the Landau gauge is
analysed within Zwanziger's local formulation of the Gribov
horizon.
\end{abstract}

\renewcommand{\theequation}{\thesection.\arabic{equation}}

\section{The model}
 In a series of papers \cite{Zwanziger:mf,Zwanziger:1992qr}, D. Zwanziger has shown that the
 restriction of the domain of integration in the path integral to
 the Gribov region $\Omega=\big\{A^a_\mu\;|\;\partial A^a=0,
 \;{{M}}^{ab}>0)\big\}$, where
${{M}}^{ab}=-\partial_\mu(\partial_\mu\delta^{ab}+gf^{acb}A^c_\mu)$
is the Faddeev-Popov operator, can be implemented by adding to the
Yang-Mills action the nonlocal horizon term
\begin{equation}
S_h=g^2\gamma^4\int d^4x
f^{abc}A^b_{\mu}({M}^{-1})^{ad}f^{dec}A^{e}_{\mu} \;.
\label{horizon}
\end{equation}
The parameter $\gamma$ is known as the Gribov parameter
\cite{Gribov:1977wm}, and is determined by the horizon condition
\cite{Zwanziger:mf,Zwanziger:1992qr}, $\frac{\delta \Gamma}{\delta
\gamma}=0$, $\Gamma$ being the quantum effective action. The
nonlocal term $\left(\mathrm{{\ref{horizon}}}\right)$ can be
localized by introducing a suitable set of additional fields
\cite{Zwanziger:mf,Zwanziger:1992qr}. The resulting action
displays two remarkable properties, namely: locality and
multiplicative renormalizability
\cite{Zwanziger:mf,Zwanziger:1992qr,Maggiore:1993wq}. Moreover,
these properties are preserved when the local composite operator
$A^{a}_{\mu}A^{a}_{\mu}$ is introduced in the theory. This enables
us to discuss the condensate $\langle{A_\mu^aA_\mu^a}\rangle$
\cite{Verschelde:2001ia,Dudal:2003vv} and the related dynamical
gluon mass $m$ in the presence of the Gribov horizon $\partial
\Omega$, within a local renormalizable framework. We give here a
sketchy account of this analysis by limiting ourselves to consider
the Gribov approximation for the horizon action
$\left(\mathrm{{\ref{horizon}}}\right)$, by setting
${M}^{ab}\approx\delta^{ab}\partial^2$. A more complete and
detailed analysis is in preparation \cite{w}. The BRST invariant
local action implementing the restriction to the region $\Omega$,
and allowing for the inclusion of the operator $A^a_\mu{A}^a_\mu$,
is $S=(S_{YM}+S_{gf}+S_h+S_{\gamma}+S_{mass})$. The term
$(S_{YM}+S_{gf})$ is the Yang-Mills action together with the
Landau gauge fixing, while $S_h$ is the localized version of the
horizon action $\left(\mathrm{{\ref{horizon}}}\right)$ in the
Gribov approximation, containing the additional fields
$\{\bar{\varphi}^a_\mu,\varphi^a_\mu\}$ and
$\{\bar{\omega}^a_\mu,\omega^a_\mu\}$. We have
\[
S_{YM}=\frac{1}{4}\int d^4x F^a_{\mu\nu} F^a_{\mu\nu}
\;,\;\;\;\;\; S_{gf}=s\int d^4x\; \bar{c}^a\partial_\mu A_\mu^a
\;,\;\;\;\;\; S_h=-s\int d^4x
\;\bar{\omega}^a_\mu\partial^2\varphi^a_\mu \;,
\]
Following \cite{Zwanziger:mf,Zwanziger:1992qr}, the term
$S_\gamma$ defines the composite operator $A^a_{\mu}\varphi^a_\mu$
and its BRST variation, introduced here through the corresponding
sources $J, \bar\lambda$. Finally, $S_{mass}$ accounts for the
mass operator $A^a_\mu{A}^a_\mu$ and its BRST variation, coupled
to the sources $\tau, \eta$.
\begin{equation}
S_\gamma=s\int{d^4x}\;\Big(\bar{\lambda}A^a_\mu\varphi^a_\mu+JA^a_\mu\bar{\omega}^a_\mu+\xi\bar{\lambda}J\Big)
\;,\hspace{0.25in}S_{mass}=\frac{1}{2}s\int{d^4x}\;\bigg(\tau
A^a_\mu{A}^a_\mu-\zeta\tau\eta\bigg)\;.
\end{equation}
The parameters $\xi$ and $\zeta$ are needed to account for the
divergences arising in the vacuum correlation functions of these
composite operators. The nilpotent BRST transformations of the
fields and sources are as follows:
\begin{eqnarray}
sA_{\mu }^{a} &=&-\left( \partial _{\mu }c^{a}+gf^{abc}A_{\mu
}^{b}c^{c}\right)\;,\;\;\;s\bar{\omega}^a_\mu=\bar{\varphi}^a_\mu\;,\;\;\;\;\;\;\;\;\;
s\bar{\lambda}=\bar{J}\;,\;\;\;\;\;\;\;\;\;\;\;\;\;\;s\tau=\eta\;,\nonumber\\
sc^{a}
&=&\frac{1}{2}gf^{abc}c^{b}c^{c}\;,\;\;\;\;\;\;\;\;\;\;\;\;\;\;\;\;\;
\;\;\;\;\;\;\;\;s\bar{\varphi}^a_\mu=0\;,\;\;\;\;\;\;\;\;\;\;\;\;
s\bar{J}=0\;,\;\;\;\;\;\;\;\;\;\;\;\;\;s\eta=0\;,\nonumber \\
s\overline{c}^{a}&=&B^{a}\;,\;\;\;\;\;\;\;\;\;\;\;\;\;\;\;\;\;\;\;\;\;
\;\;\;\;\;\;\;\;\;\;\;\;\;\;\;\;\;\;\;\;s\varphi^a_\mu=\omega^a_\mu\;,\;\;\;\;\;\;\;\;
s\lambda=0\;,\nonumber \\
sB^{a}
&=&0\;,\;\;\;\;\;\;\;\;\;\;\;\;\;\;\;\;\;\;\;\;\;\;\;\;\;\;\;\;\;\;\;
\;\;\;\;\;\;\;\;\;\;\;\;s\omega^a_\mu=0\;,\;\;\;\;\;\;\;\;\;\;\;\;sJ=\lambda\;.
\end{eqnarray}
By making use of the algebraic renormalization \cite{Piguet:er},
the action $S=S_{YM}+S_{gf}+S_h+S_{\gamma}+S_{mass}$ turns out to
be multiplicatively renormalizable. In particular, as discussed in
\cite{Zwanziger:mf,Zwanziger:1992qr}, the horizon condition is
obtained by setting the sources $(J,{\bar J}, \lambda,
\bar{\lambda})$ equal to $J={\bar J}=\gamma^2$, $\lambda={\bar
\lambda}=0$, and by requiring that $\frac{\delta \Gamma}{\delta
\gamma}=0$.

\section{Gap equation and propagators}

Proceeding as in \cite{Sobreiro:2004us}, it is not difficult to
evaluate the effective action $\Gamma(\gamma)$ at one-loop level,
in the presence of the dynamical gluon mass $m$. The horizon
condition  $\frac{\delta \Gamma}{\delta \gamma}=0$ leads to the
following gap equation
\begin{equation}
\frac{3Ng^2}{4}\int\;\frac{d^4k}{(2\pi)^4}\frac{1}{k^4+m^2k^2+\gamma^4}=1\;,\label{gap}
\end{equation}
which generalizes that obtained in
\cite{Gribov:1977wm,Zwanziger:mf,Zwanziger:1992qr}. Notice now
that the dynamical mass $m$ appears explicitly in eq.
$\left(\mathrm{{\ref{gap}}}\right)$. The gap equation
$\left(\mathrm{{\ref{gap}}}\right)$ can be used to obtain the
gluon and ghost propagators in the tree-level approximation. The
gluon propagator is found to be \cite{Sobreiro:2004us}
\begin{equation}
\langle{A_\mu^a(q)A_\nu^b(-q)}\rangle=\delta^{ab}\frac{q^2}{q^4+m^2q^2+\gamma^4}
\Bigg(\delta_{\mu\nu}-\frac{q_\mu{q}_\nu}{q^2}\Bigg)\;.\label{glu}
\end{equation}
For the ghost two point function we have \cite{Sobreiro:2004us}
\begin{equation}
G(q)=\langle{c^a(q)\bar{c}^a(-q)}\rangle
\sim\frac{1}{q^4}\;.\label{gho}
\end{equation}
Notice that, according to the
\cite{Gribov:1977wm,Zwanziger:mf,Zwanziger:1992qr} the gluon
propagator is suppressed in the infrared, while the ghost
propagator is enhanced.

\section{Conclusions}

The restriction of the domain of integration to the region
$\Omega$ has been implemented by considering Zwanziger's local
action  \cite{Gribov:1977wm,Zwanziger:mf,Zwanziger:1992qr} in the
Gribov approximation. Expression
$\left(\mathrm{{\ref{gap}}}\right)$ is the gap equation defining
the Gribov parameter $\gamma$ in the presence of the dynamical
gluon mass $m$. The resulting gluon propagator is suppressed in
the infrared, while the ghost propagator is enhanced. This
behavior of the gluon and ghost propagators is in agreement with
that found in \cite{Gribov:1977wm,Zwanziger:mf,Zwanziger:1992qr}.
It has also been found in \cite{Alkofer:2000wg} within the
Schwinger-Dyson framework. Evidences for a dynamical gluon mass in
the Landau gauge within the Schwinger-Dyson formalism have been
obtained recently in \cite{Aguilar:2004kt}. Finally, lattice
simulations have provided confirmations of the infrared
suppression of the gluon propagator and of the ghost enhancement,
see \cite{Bloch:2003sk} and refs. therein, reporting a gluon mass
$m$ of the order of $\approx$ 600 $MeV$ \cite{Langfeld:2001cz}.

\section{Acknowledgments}

The CNPq-Brazil, the SR2-UERJ and the CAPES-Brazil are gratefully
acknowledged for financial support.

\end{document}